\documentclass[aps,prb,twocolumn]{revtex4}
\usepackage{amsfonts}
\usepackage{hyperref}
\usepackage{amsmath}
\usepackage{amsthm}
\usepackage{graphics}
\usepackage{graphicx}
\usepackage{subfigure}
\usepackage{amssymb}
\usepackage{graphics}
\usepackage{graphicx}
\usepackage{epstopdf}
\usepackage{color}
\usepackage{subfigure}
\usepackage{pgfplots,tikz}
\usepackage{url}
\usepackage{float}
\usepackage{soul}
\usepackage{booktabs}
\newcommand\dertt[1]{ \frac{\partial{ #1}}{\partial t} }

\def\Mp{M_{\rm p}}
\def\Rp{a_{\rm p}}

\def\Vp{V_\mathrm{p}}
\def\x{\mathbf{x}}
\def\q{\mathbf{q}}

\begin{document}

\title{Quantum vortex reconnections mediated by trapped particles}
\author{Umberto Giuriato}
\affiliation{
Universit\'e C\^ote d'Azur, Observatoire de la C\^ote d'Azur, CNRS, Laboratoire Lagrange, Nice, France}
\author{Giorgio Krstulovic}
\affiliation{
Universit\'e C\^ote d'Azur, Observatoire de la C\^ote d'Azur, CNRS, Laboratoire Lagrange, Nice, France}

\begin{abstract}
Reconnections between quantum vortex filaments in presence of trapped particles are investigated using numerical simulations of the Gross--Pitaevskii equation. 
Particles are described with classical degrees of freedom and modeled as highly repulsive potentials which deplete the superfluid. First, the case of a vortex dipole with a single particle trapped inside one of the vortices is studied. It is shown that the reconnection takes place at the position of the particle as a consequence of the symmetry breaking induced by it. The separation rate between the reconnecting points is compatible with the known dynamics of quantum vortex reconnections and it is independent of the particle mass and size. After the reconnection, the particle is pushed away with a constant velocity and its trajectory is deflected because of the transverse momentum exchange with the vortex filaments. The momentum exchanges between the particle, the vortex, and a density pulse are characterized. Finally, the reconnection of two linked rings, each of them with several initially randomly distributed particles is studied. It is observed that generically, reconnections take place at the location of trapped particles. It is shown that reconnection dynamics is unaffected for light particles.
\end{abstract}
\maketitle

\section{Introduction}
One of the most striking feature of superfluids is the presence of quantum vortices, thin tornadoes which arise as topological defects and nodal lines of the complex order parameter describing the system \cite{donnellyQuantizedVorticesHelium1991}. Quantum vortices have been observed in different kind of superfluids: from atomic Bose--Einstein condensates (BECs), where their core is micrometer sized to
superfluid  $^4$He, where the core size is of a few Angstroms.
The topological nature of quantum vortices constrains their circulation to be a discrete multiple of the quantum of circulation $\Gamma=h/m$, where $h$ is the Planck constant and $m$ is the 
mass of the bosons constituting the superfluid. 

The dynamics of such vortex filaments is rich and still not fully comprehended. 
In particular, a fundamental phenomenon is the occurrence of
reconnection events. In general, in fluid 
mechanics a vortex reconnection is an event in which the topology of the vorticity field is rearranged \cite{kidaVortexReconnection1994}.
In the case of classical fluids, the presence of viscosity breaks the Kelvin circulation theorem, allowing the reconnection between vortex tubes \cite{hussainMechanicsViscousVortex2011}.
In the case of inviscid superfluids, the vorticity is supported exclusively along the unidimensional vortex filaments and the reconnection between them is made possible because of the vanishing density at the core of the vortices
\cite{koplikVortexReconnectionSuperfluid1993}.
Specifically, the process of superfluid vortex reconnection consists in the local exchange of two strands of different filaments after a fast approach, allowing the topology to vary.
In quantum turbulence, reconnections are also thought to be a fundamental mechanism for the redistribution of energy at scales smaller than the inter-vortex distance \cite{vinenQuantumTurbulence2002}.

The separation $\delta(t)$ between the two reconnecting points is the simplest observable 
that characterizes a vortex reconnection.
Given that a reconnection is an event localized in space and time, sufficiently close to 
the reconnection event
it is expected to be fully driven 
by the interaction between two filaments. Assuming that at this scale the only parameter that determines the dynamics is the circulation $\Gamma$ about each filament, a simple dimensional analysis suggests the following scaling for the separation rate:
\begin{equation}
\delta(t)=A^{\pm}\left(\Gamma \left|t - t_\mathrm{rec}\right|\right)^{1/2},
\label{Eq:delta}
\end{equation}
where $A^{\pm}$ are dimensionless pre-factors, $t_\mathrm{rec}$ is the reconnection time and the labels $-$ and $+$ refer respectively to before and after the reconnection event.
Such scaling has been demonstrated analytically in the context of the Gross--Pitaevskii (GP) model for $\delta\to0$ \cite{nazarenkoAnalyticalSolutionNonlinear2003,villoisUniversalNonuniversalAspects2017,promentMatchingTheoryCharacterize2020},
and it has been observed to be valid even at distances that go beyond several healing lengths \cite{villoisUniversalNonuniversalAspects2017,galantucciCrossoverInteractionDriven2019}. Note that previous studies reported disparate exponents that still need to be explained \cite{zuccherQuantumVortexReconnections2012, allenVortexReconnectionsAtomic2014, roraiApproachSeparationQuantised2016b}. The scaling (\ref{Eq:delta}) has been also observed Biot-Savart simulations \cite{galantucciCrossoverInteractionDriven2019,tsubotaSimulationCounterflowTurbulence2011,baggaleyThermallyMechanicallyDriven2012} and superfluid helium experiments \cite{paolettiReconnectionDynamicsQuantized2010}. If an external driving mechanisms is absent, the scaling \eqref{Eq:delta} is considered an universal feature of vortex reconnections, as well as the fact that the filaments always approach slower than they separate, i.e. $A^+/A^->1$. This last observation has been explained by a novel matching theory as the consequence of an irreversible mechanism related to the sound radiated during the event \cite{villoisIrreversibleDynamicsVortex2020,promentMatchingTheoryCharacterize2020}. 

In recent years, vortex reconnections have been directly observed in atomic BECs, by means of destructive absorption imaging \cite{serafiniDynamicsInteractionVortex2015} and in superfluid helium experiments by using solidified 
hydrogen particles as probes
\cite{bewleyCharacterizationReconnectingVortices2008,paolettiReconnectionDynamicsQuantized2010}.
This latter technique has become a standard tool for the investigation of the properties of superfluid helium and quantum vortices, since its first utilization in 2006
\cite{bewleyVisualizationQuantizedVortices2006}. Indeed, such particles get captured by quantum vortices thanks to pressure gradients and are carried by them, unveiling in this way the dynamics of
the filaments. Besides the reconnections between vortices and Kelvin waves (helicoidal displacements that propagate along the vortex filaments), solidified hydrogen particles succeeded in revealing
important differences between the statistics of classical and quantum turbulent states \cite{mantiaQuantumClassicalTurbulence2014,lamantiaQuantumTurbulenceVisualized2014a}.
However, being the typical size of such particles four order of magnitude larger than the vortex core size, it is far from being trivial that they actually behave as tracers. For this reason, understanding the 
actual vortex-particle interactions and how particles and fluids affect the motions of each other is a crucial theoretical task.

Many models have been developed and studied in this regard. The main difficulty is caused by the large extent of the scales involved in the problem, so that different phenomenological approaches need to be used. For what concerns the large scales, the dynamics of particles in classical fluids has been phenomenologically adapted to the two-fluid description of a superfluid
 \cite{pooleMotionTracerParticles2005} and the distribution of inertial passive particles has been studied in the Hall-Vinen-Bekarevich-Khalatnikov model \cite{polancoInhomogeneousDistributionParticles2020}. 
 In this macroscopic approach, the vorticity is a coarse-grained field and there is no notion of quantized vortices. Instead in the vortex-filament model, the superfluid is modeled as a collection of filaments that evolve according to Biot-Savart integrals \cite{donnellyQuantizedVorticesHelium1991}. 
 This method involves non-local contributions and a singular integral for the computation of the vortex self-induced velocity that needs to be regularized \cite{schwarzThreedimensionalVortexDynamics1988}.
In this framework, hard spherical particles can be modeled as moving boundary conditions \cite{kivotidesInteractionsParticlesQuantized2008,sergeevParticlesVortexInteractionsFlow2009}, although both the reconnections between vortices as well as between a vortex and a particle surface need to be implemented with an ad-hoc procedure.
 These issues are absent in the GP model, in which the evolution of the order parameter of the superfluid is described with a nonlinear Schr$\mathrm{\ddot{o}}$dinger equation. 
 Indeed, although it is formally derived for dilute Bose-Einstein condensates, the GP model can be considered as a general prototype of a superfluid at very low temperature, including superfluid helium. Unlike the vortex-filament method or the HVBK model, the full dynamics of vortices emerges naturally, including the reconnection events. Particles modeled as highly repulsive potentials have been successfully implemented in the GP framework, allowing for an extensive study of the capture process \cite{giuriatoInteractionActiveParticles2019}, the interaction between trapped particles and Kelvin waves \cite{giuriatoHowTrappedParticles2020a}, and the Lagrangian properties of quantum turbulence \cite{giuriatoActiveFinitesizeParticles2020a}. Recently, the dynamics of particles trapped inside GP vortices has been addressed also in two 
 dimensions \cite{griffinMagnusforceModelActive2020}.

Being the GP equation a microscopic model, regular at the vortex core, it is the natural setting where quantum vortex reconnections can be studied. In this work, we combine such suitability
with the simplicity of modeling particles in the GP framework, to study vortex reconnections in presence of particles trapped by the filaments.
We focus on two different configurations. In 
section \ref{sec:dipole} we study the evolution of
a dipole of two counter-rotating straight vortices with a particle trapped in one of them. In section \ref{sec:rings} we characterize the reconnection of two linked rings loaded with a number of particles. In the 
first case the reconnection is induced by the presence of the particle, and its simplicity allows for a systematic investigation of the mutual interaction between vortices and {particles} during the process of the reconnection. In the second case, the reconnection happens even in absence of particles, so that how the presence of particles effectively affect the reconnection process can be addressed. 

\section{Model for particles and quantum vortices}
We consider a quantum fluid with $N_\mathrm{p}$ spherical particles of mass $\Mp$ and radius $a_\mathrm{p}$ immersed in it. We describe the system by a self-consistent model based on the three-dimensional Gross-Pitaevskii equation. The particles are modeled by strong localized potentials $\Vp$, that completely deplete the superfluid up to a distance $a_\mathrm{p}$ from the position of their center $\mathbf{q}_i$.
The dynamics of the system is governed by the following Hamiltonian:
\begin{eqnarray}
H&=&\int\left(\frac{\hbar^2}{2m}|\nabla\psi|^2-\mu|\psi|^2+\frac{g}{2}|\psi|^4 + {\sum_{i=1}^{N_\mathrm{p}}V_\mathrm{p}(\mathbf{r}-\mathbf{q}_i)|\psi|^2} \right)\mathrm{d}\mathbf{r}\nonumber\\
&+& {\sum_{i=1}^{N_\mathrm{p}}\frac{\left(\mathbf{p}^\mathrm{part}_i\right)^2}{2M_\mathrm{p}}} 
+ {\sum_{i<j}^{N_\mathrm{p}}V_\mathrm{rep}^{ij}}.
\label{Eq:Hamiltonian}
\end{eqnarray}
where $\psi$ is the order parameter of the quantum and $\mathbf{p}_i^\mathrm{part}=M_\mathrm{p}\mathbf{\dot{q}}_i$ are the particles linear momenta. The chemical potential is denoted by $\mu$.
The nonlinear self-interaction coupling constant of the fluid is denoted by $g$ and $m$ is the mass of the condensed bosons.  The potential  $V_\mathrm{rep}^{ij}$ is a repulsive potential between particles, needed to avoid an unphysical overlap, due to a short range fluid mediated interaction \cite{shuklaStickingTransitionMinimal2016,giuriatoClusteringPhaseTransitions2019}.
The equations of motion for the superfluid field $\psi$ and the particle positions $\q_i=(q_{i,x},q_{i,y},q_{i,z})$ are:
\begin{eqnarray}
i\hbar\dertt{\psi} = - \frac{\hbar^2}{2m}\nabla^2 \psi +  g|\psi|^2\psi-\mu\psi+\sum_{i=1}^{N_\mathrm{p}}\Vp(| \x -{\bf q}_i |)\psi \label{Eq:GPEParticles}, \nonumber \\
\\
\Mp\ddot{\bf q}_i = - \int  \Vp(| \x -{\bf q}_i|) \nabla|\psi|^2\, \mathrm{d} \x+\sum_{j\neq i}^{N_\mathrm{p}}\frac{\partial}{\partial{\bf q}_i }V_\mathrm{rep}^{ij}. \nonumber \\ 
\label{Eq:GP}
\end{eqnarray}
We refer to \cite{shuklaParticlesFieldsSuperfluids2018,giuriatoInteractionActiveParticles2019,giuriatoHowTrappedParticles2020a,giuriatoActiveFinitesizeParticles2020a} for further details about the model, where it has been recently adopted to study extensively the interaction between particles and quantum vortices.

In absence of particles, the ground state of the system is a homogeneous flat condensate $\psi_\infty=\sqrt{\mu/g}\equiv\sqrt{{\rho_\infty}/{m}}$, with a constant mass density $\rho_\infty$.  Linearizing around this value, dispersive effects take place at scales smaller than the healing length $\xi=\sqrt{\hbar^2/2g\rho_\infty }$, while large wavelength excitations propagate with the phonon (sound) velocity $c=\sqrt{g\rho_\infty/m^2}$.
The close relation between the GP model and hydrodynamics comes from the Madelung transformation $\psi(\x)=\sqrt{{\rho(\x)}/{m}}\,e^{i\frac{m}{\hbar}\phi(\x)}$, that maps the GP \eqref{Eq:GPEParticles} into the continuity and Bernoulli equations of a superfluid of density $\rho$ and velocity $\mathbf{v}_\mathrm{s}=\nabla\phi$. Although the superfluid velocity is potential, the phase is not defined at the nodal lines of $\psi(\x)$ and thus vortices may appear as topological defects. Each superfluid vortex carries a quantum of circulation 
$\Gamma = h/m = 2\pi \sqrt{2} c\xi$, and they are characterized by a vanishing density core size of the order of $\xi$. 

In this work, we perform numerical simulations of the coupled differential equations (\ref{Eq:GPEParticles},\ref{Eq:GP}) in a periodic cubic box of side $L=128\xi$ with $N_{\mathrm{c}}=256^3$ collocation points. We use a standard pseudospectral method with a fourth-order Runge-Kutta scheme for the time-stepping. In numerics, we measure distances in units of $\xi$, velocities in units of
$c$ and times in units of $\tau=\xi/c$. 
As described in the Appendix \ref{dealiasing} and in Ref. \cite{krstulovicEnergyCascadeSmallscale2011}, {de-aliasing is applied to equations (\ref{Eq:GPEParticles},\ref{Eq:GP}), in such a way that} they conserve the total energy $H$ \eqref{Eq:Hamiltonian}, the total fluid mass $N=\int |\psi|^2\,\mathrm{d}\mathbf{x}$ and the total momentum 
\begin{equation}
\mathbf{p}^\mathrm{tot} = \mathbf{p}^\mathrm{GP}+\sum_{i=1}^{N_\mathrm{p}}\mathbf{p}^\mathrm{part},
\label{Eq:momcons}
\end{equation}
where $\mathbf{p}^\mathrm{GP}=i\hbar/2\int \left( \psi\nabla\psi^*-\psi^*\nabla\psi \right)\,\mathrm{d}\mathbf{x}$ is the momentum of the quantum fluid. If dealising is not carefully performed, the discrete system does not conserved momentum. In the simulations presented here the total momentum is conserved up to $8$ decimal digits.

We use two different particle potentials to model the particles. For the simulations with the dipole, a smoothed hat-function $\Vp^1(r)=\frac{V_0}{2}(1-\tanh\left[\frac{r^2 -\zeta^2}{4\Delta_a^2}\right])$ is used. The parameters $\zeta$ and $\Delta_a$ are set to model the particle attributes. In particular, $\zeta$ fixes the width of the potential and it is related to the particle size, while $\Delta_a$ controls the steepness of the smoothed hat-function. The latter needs to be adjusted in order to avoid Gibbs effect in the Fourier transform of $V_\mathrm{p}^1$. For the simulations of the Hopf link, we use a Gaussian potential $\Vp^2(r)=V_0\exp{(-r^2/2 d_{\rm eff}^2})$, where the width is fixed using the Thomas-Fermi approximation to set an approximate radius $\zeta$ to the particle as $d_{\rm eff}=\zeta/\sqrt{2\log(V_0/\mu)}$.
Since the particle boundaries are not sharp, the effective particle radius is measured as $\Rp=(3M_0/4\pi\rho_\infty)^\frac{1}{3}$, where $M_0=\rho_\infty L^3(1-\int |\psi_\mathrm{p}|^2\,\mathrm{d}\mathbf{x}/\int |\psi_\infty|^2\,\mathrm{d}\mathbf{x})$ is the fluid mass displaced by the particle and $\psi_\mathrm{p}$ is the steady state with just one particle. Practically, given the set of numerical parameters $\zeta$ and $\Delta_a$, the state $\psi_\mathrm{p}$ is obtained numerically with imaginary time evolution and the excluded mass $M_0$ is measured directly.  We use the repulsive potential $V_\mathrm{rep}^{ij}=\gamma_\mathrm{rep}(2\Rp/|{\bf q}_i-{\bf q}_j|)^{12}$ in order to avoid an overlap between them. The functional form of $V_\mathrm{rep}^{ij}$ is inspired by the repulsive term of the Lennard-Jones potential and the pre-factor $\gamma_\mathrm{rep}$ is adjusted numerically so that  the inter-particle distance $2\Rp$ minimizes the sum of $V_\mathrm{rep}^{ij}$ with the fluid mediated attractive potential \cite{shuklaStickingTransitionMinimal2016,giuriatoClusteringPhaseTransitions2019}.

The initial condition for the dipole and a single ring (without particles) are obtained using a Newton--Raphson method and a biconjugate-gradient technique in order to minimize the sound emission \cite{abidGrossPitaevskiiDynamics2003}. 
The Hopf link of two rings is obtained multiplying two states containing a ring each. 

\section{Reconnection of a vortex dipole}
\label{sec:dipole}
We start by presenting a series of numerical simulations of a dipole of two counter-rotating superfluid vortices, with a single particle initially trapped inside one of them.
Such setting is useful to illustrate how a superfluid vortex reconnection can be triggered by the symmetry breaking produced by the presence of particles. Indeed, in absence of trapped particles, the vortex dipole is a steady configuration, in which a spontaneous self-reconnection does not happen unless a Crow instability is induced \cite{berloffMotionBoseCondensate2001}.
At the same time, the simplicity of the initial configuration allows for a systematic study of the mutual effects between the particle and the reconnecting filaments. 

In the initial time of each simulation, the vortices are straight and aligned along the $z$ direction. The initial velocity of the particle is set equal to the translational speed of the dipole 
$\mathbf{v}_\mathrm{d} \sim \left(\Gamma/2\pi d\right)\mathbf{\hat{y}}$,
where $d$ is the distance between the two filaments and $\mathbf{\hat{y}}$ is the unit vector along the $y$ direction 
\cite{griffinMagnusforceModelActive2020,saffmanVortexDynamics1992}.
We performed the same experiment using particles of two different sizes and for a wide range of mass densities. 

It has been observed in Ref. \cite{giuriatoInteractionActiveParticles2019} that the effective Hamiltonian describing the process of particle capture by 
a vortex induces a dynamics which is invariant 
under the following scaling transformation:
\begin{equation}
d\rightarrow \lambda d,\quad a_\mathrm{p}\rightarrow \lambda a_\mathrm{p},\quad t\rightarrow \lambda^2 t\qquad \forall \lambda\in\mathbb{R}^+,
\label{Eq:scaling}
\end{equation}
where $d$ is the vortex-particle distance.
In order to check if the scaling invariance \eqref{Eq:scaling} is valid also in the present simulations, we choose the radius of the large particle exactly $\lambda=2$ times larger than the radius of the small one. Analogously, in the case of the large particle, the vortex filaments are initially placed $\lambda=2$ times more distant than for the small particle. 
If such invariance subsists, it would be an indication of the analogy between the reconnection process and the trapping mechanism. Besides, it would naturally extend the validity of the results reported below in the case of particles with larger sizes, comparable with the ones used in current experiments. Note however that the scaling invariance \eqref{Eq:scaling} 
neglects the density profile of the vortex core, as well as other more complex particle-vortex interactions which can become relevant when a particle is trapped, like the Magnus effect.

The parameters used for these sets of simulations summarized in Table \ref{tab:dipole_parameters} (note that the repulsive potential $V_\mathrm{rep}^{ij}$ in Eq. (\ref{Eq:GP}) is absent because only one particle is present).
\begin{table}[htb]\
  \caption{%
    Simulation parameters for the vortex dipole reconnection experiment.}
    \label{tab:dipole_parameters}
  \newcommand*{\ct}[1]{\multicolumn{1}{c}{#1}}  
  \begin{ruledtabular}  
  \begin{tabular}{cccccc}
	  $\lambda$ & $d/\xi$ & $a_\mathrm{p}/\xi$ & $\quad\zeta/\xi\quad$ & $\quad\Delta_a/\xi\quad$ & $\quad V_0/\mu$ \\
    \midrule                   
    $1$ & $10$ & $4.3$ & $3.0$ & $0.75$ & $20$ \\
    $2$ & $20$ & $8.6$ & $7.4$ & $0.75$ & $20$ \\ 
  \end{tabular}
  \end{ruledtabular}
\end{table}

\begin{figure*}
\includegraphics[width=.99\linewidth]{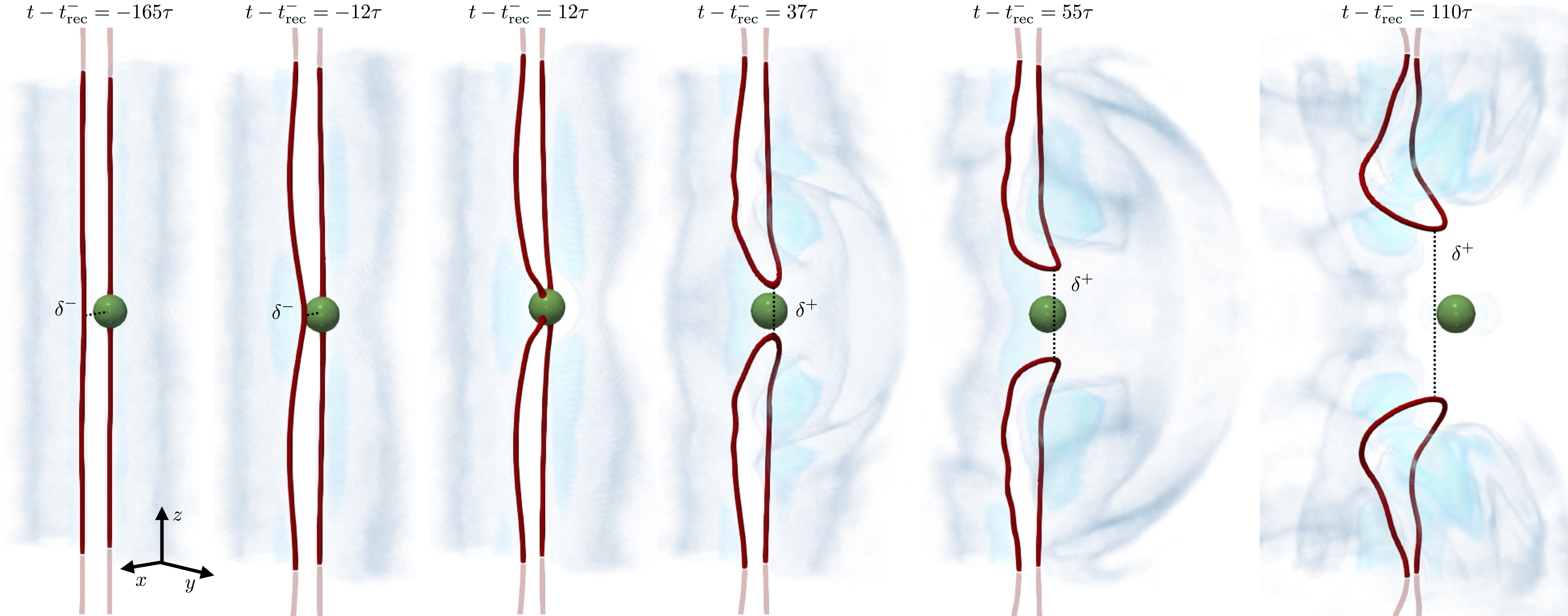}
\caption{(\textit{Color online}) Snapshots of the superfluid density and a neutral-mass particle of size $a_\mathrm{p}=4.3\xi$ during the dipole reconnection (time varies from left to to right). The initial distance between the vortices is $d=10\xi$. Vortices are displayed as red isosurfaces at low density, particles are the green spheres and sound is rendered in blue.}
\label{Fig:dipolerec}
\end{figure*}

Snapshots of the typical evolution of the dipole configuration under the GP dynamics (\ref{Eq:GPEParticles},\ref{Eq:GP}) are displayed in Fig.\ref{Fig:dipolerec}, for a neutral particle of size $a_\mathrm{p}=4.3\xi$ and initial vortex separation $d=10\xi$.
During the motion of the dipole, the particle starts to precede about the filament because of Magnus effect \cite{kiknadzeMagnusKuttaJukovskiiForce2006,giuriatoHowTrappedParticles2020a,griffinMagnusforceModelActive2020}. At the same time, the two vortices start to bend,
until the filament without particle reconnects with the surface of the sphere at time $t_\mathrm{rec}^-$. After the reconnection, the contact point of the free vortex separates into two branches, which then slide on the particle surface towards opposite directions. For a time window of about $\sim 20\tau$ the particle is pierced by both vortices, until the couple of pinning points above and below the particles merge and the vortices detach symmetrically. The reconnection changes the topology of the flow, so that the dipole is eventually converted into a single vortex ring (that in Fig.\ref{Fig:dipolerec} appears folded on the vertical direction because of spatial periodicity). At the time of the detachment, a clear spherical sound pulse is generated at the reconnection point. It expands and propagates along the $y$ direction, which is the dipole propagation direction and coincides to the normal to the reconnection plane, in agreement with references \cite{villoisIrreversibleDynamicsVortex2020,promentMatchingTheoryCharacterize2020}. Simultaneously, the particle is realeased and abruptly accelerated. Eventually, it keeps on moving forward with a constant speed larger than the dipole velocity.

\begin{figure}[h!]
\includegraphics[width=.99\linewidth]{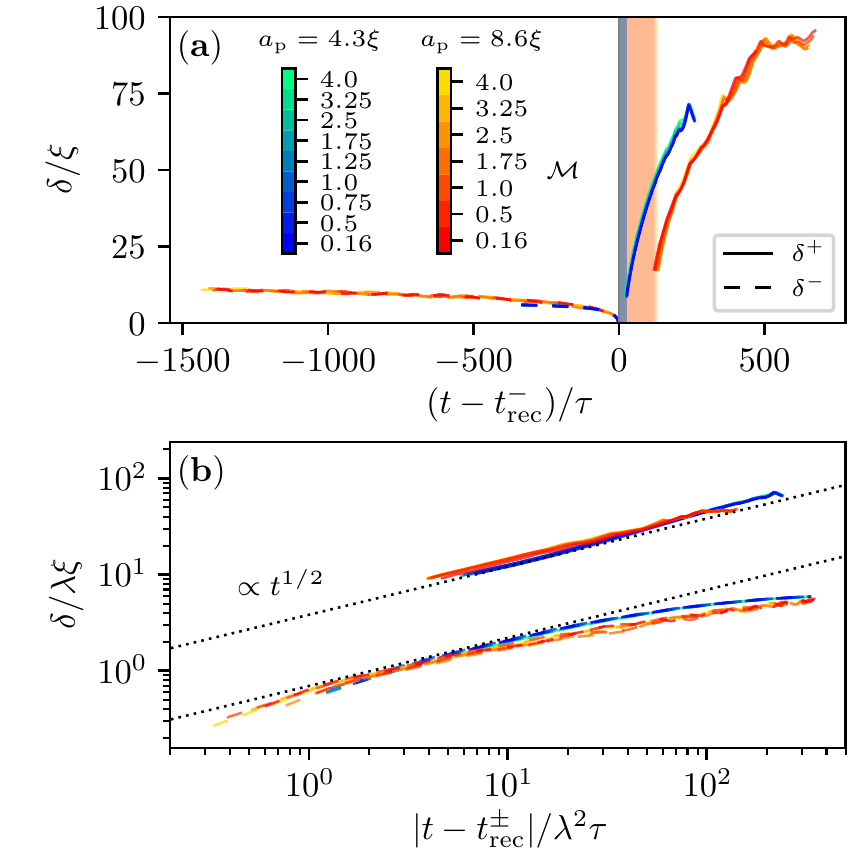}
\caption{(\textit{Color online}) \textbf{(a)} Distance $\delta(t)$ between reconnecting points for particles of size $a_\mathrm{p}=4.3\xi$ and $a_\mathrm{p}=8.6\xi$. Dashed lines correspond to $\delta^-$ before reconnection and solid lines correspond to $\delta^+$
after reconnection. \textbf{(b)} Log-Log plot of $\delta(t)$, with the rescaling \eqref{Eq:scaling}. $\lambda=1$ for the particle of size $a_\mathrm{p}=4.3\xi$ and $\lambda=2$ for the particle of size $a_\mathrm{p}=8.6\xi$. Dotted lines indicate the scaling $t^{1/2}$.}
\label{Fig:delta}
\end{figure}

Before exploring more in detail the origin of the particle dynamics, we address the question whether the observed reconnections induced by the particle are compatible with the standard picture of GP reconnections. In order to do so, we compute the separation $\delta(t)$ between the 
reconnecting points as a function of time. When the circulation $\Gamma$ is the only relevant parameter driving the reconnection dynamics, $\delta(t)$ is expected to scale as Eq. \eqref{Eq:delta}.
We operatively define the separation before the reconnection $\delta^-$ as the distance between the reconnecting point on the free vortex and the particle surface. After the reconnection time $t_\mathrm{rec}^-$ between the free vortex and the sphere surface, the separation is not well defined until the particle detachment, after which $\delta^+$
becomes simply the distance between the two extremal points of the outgoing vortex ring (see Fig.\ref{Fig:dipolerec}). The vortex filaments have been tracked using the method based on the {pseudo-vorticity} developed in 
\cite{villoisVortexFilamentTracking2016}.
Since the initial measurable value of $\delta^+$ is of the order of the
particle diameter $2a_\mathrm{p}$, we extrapolate the virtual original time time $t_\mathrm{rec}^+$ at which $\delta^+(t_\mathrm{rec}^+)=0$ performing a linear fit of $(\delta^+(t))^2$ and evaluating the point where it vanishes. The same protocol has been used with $\delta^-(t)$ to refine also the value of $t_\mathrm{rec}^-$.  The evolution of $\delta(t)$ is displayed in Fig.\ref{Fig:delta}.a for all the types of particles analyzed. 
In Fig.\ref{Fig:delta}.b, $\delta^+(t)$ and $\delta^-(t)$ are plotted in a logarithmic scale, after rescaling the distances by a factor $\lambda$ and times by a factor $\lambda^2$ ($\lambda=1$ for the small particle and $\lambda=2$ for the large one), according to Eq. (\ref{Eq:scaling}). It is apparent that the separation rate is independent of the particle mass and always shows a scaling compatible with $t^{1/2}$. This evidence confirms that, although the reconnection is triggered by the presence of the particle, the vortex dynamics is effectively fully governed only by the circulation. Moreover, the scaling invariance (\ref{Eq:scaling}) seems to be respected for the separation rate.
Finally, note that the observed positive ratio between the pre-factors of the separation rate (\ref{Eq:delta}) after and before the reconnection ($A^+/A^- \sim 5.5$) is consistent with the irreversibility of the reconnection dynamics, which is related to the conversion of energy into sound \cite{villoisUniversalNonuniversalAspects2017,villoisIrreversibleDynamicsVortex2020,promentMatchingTheoryCharacterize2020}.

\begin{figure}[h!]
\includegraphics[width=.99\linewidth]{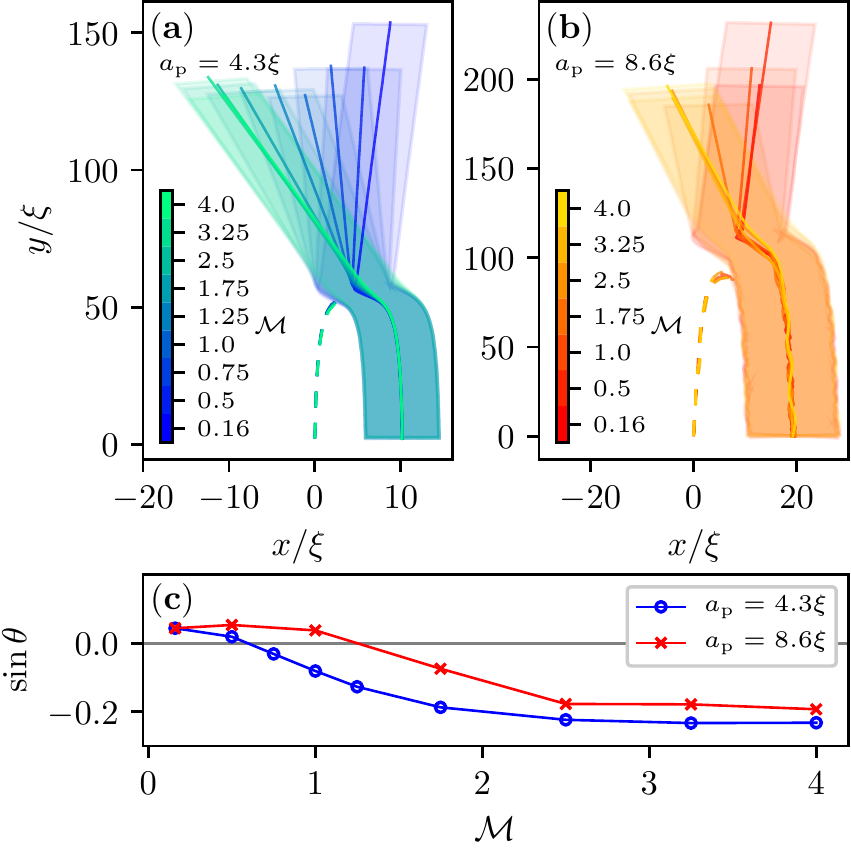}
\caption{(\textit{Color online}) In the top panel, trajectories of particles of size $a_\mathrm{p}=4.3\xi$ \textbf{(a)} and  $a_\mathrm{p}=8.6\xi$ \textbf{(b)} during the dipole reconnection. Different colors correspond to different masses and the shaded regions indicate the area spanned by each particle. The dashed lines of corresponding colors are the trajectories of the reconnecting point of the vortex without particle at times 
$t<t_\mathrm{rec}^-$. \textbf{(c)} Angle of deflection of the particle trajectory after the reconnection as a function of the particle mass, for both particle sizes (blue circles correspond to $a_{\mathrm{p}}=4.3\xi$ and red crosses to $a_{\mathrm{p}}=8.6\xi$). The angle considered is with respect to the dipole propagation direction. }
\label{Fig:traj}
\end{figure}

In Fig.\ref{Fig:traj}.a and b we show the trajectories of the particles on the plane orthogonal to the dipole, respectively for the small and the large particle and for all the different masses used. 
The shaded regions indicate the actual area spanned by each particle.
In the same figure, the dashed lines show the trajectories of the reconnecting point on the vortex without particle (initially placed at $x=0$, $y=0$) until it touches the particle surface at time $t_\mathrm{rec}^-$. For the large particle one can appreciate the different Magnus precession frequencies, inversely proportional to the mass. We observe that the ballistic motion of the particle after the reconnection is deflected with respect to the propagation direction of the dipole and a correlation between the particle mass and the deflection angle is apparent. In particular, the heaviest particles show a smooth trajectory and a deflection concordant with the velocity orientation at the reconnection point. 
Conversely, light particles are slightly bounced back in the opposite direction. 
In Fig.\ref{Fig:traj}.c the deflection angle
$\theta$ of the particle trajectory with respect to the dipole propagation direction is displayed as a function of the particle mass. As already qualitatively observed in Fig.\ref{Fig:traj}.a and b, both the small and the large particle (indicated respectively by blue circles and red crosses) deviate in a similar manner, with the deflection angle that saturates at $\sin\theta\sim -0.2$ for the largest masses.
The origin of such behavior can be understood as the consequence of a transverse momentum transfer between the vortices and the particle, that we analyze in the remanent of this section.

\begin{figure}[h!]
\includegraphics[width=.99\linewidth]{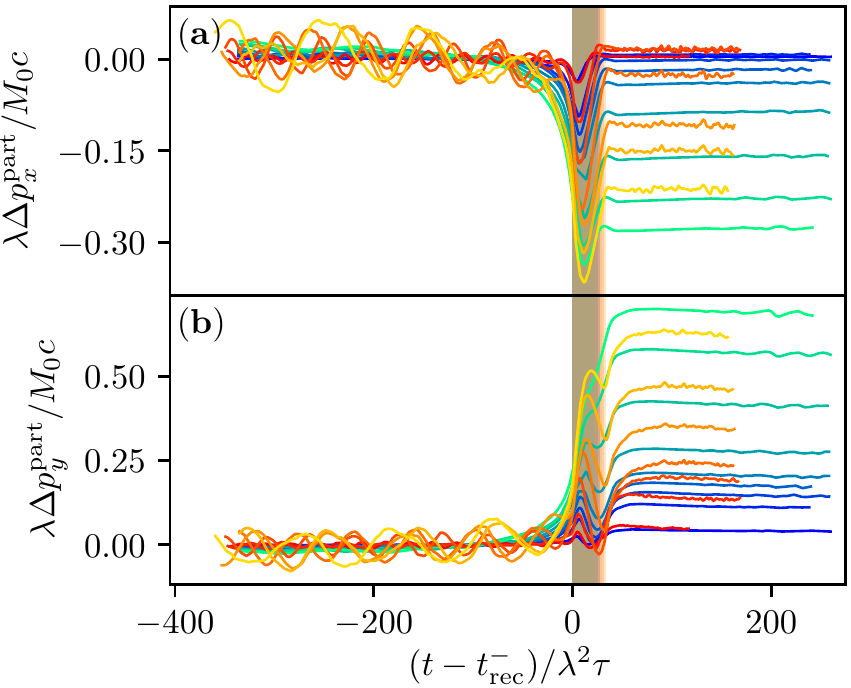}
\caption{(\textit{Color online}) $x$ component \textbf{(a)} and $y$ component \textbf{(b)} of the particle momentum increment with the rescaling (\ref{Eq:scaling}) as a function of time. Different colors correspond to different particle species, with the same convention of Fig.\ref{Fig:delta}. }
\label{Fig:momentum}
\end{figure}

The $x$ component and the $y$ component particle momentum increment $\Delta\mathbf{p}^\mathrm{part}(t) = \mathbf{p}^\mathrm{part}(t) - \mathbf{p}^\mathrm{part}(t=0)$ are plotted as a function of the rescaled time respectively in Fig.\ref{Fig:momentum}.a and b. 
The data associated with all the species of particles analyzed are displayed using the same convention of Fig.\ref{Fig:traj} and also the particle momentum has been rescaled as $\mathbf{p}^\mathrm{part}\rightarrow \mathbf{p}^\mathrm{part}/\lambda$, according to the transformation (\ref{Eq:scaling}).
Note that at the initial time of the simulations the particle is placed in the reference frame co-moving with the dipole, so that its momentum is aligned with 
the propagation direction of the dipole (the $y$ direction) and
reads $\mathbf{p}^\mathrm{part}(t=0) = M_\mathrm{p}\mathbf{v}_\mathrm{d} = \left(M_\mathrm{p}\Gamma/2\pi d\right)\mathbf{\hat{y}}$. 
We can clearly observe the 
abrupt acceleration felt by the particle in both the transverse and longitudinal directions during the reconnection event, followed after the detachment by a relaxation to a ballistic motion with constant speed. 
The ballistic motion is due to the absence of Stokes drag in the superfluid, and a negligible interaction with sound or with the outgoing vortex ring.
The shaded area represents the time window after $t_\mathrm{rec}^-$ in which the particle is pierced by both the filaments and the vortex separation $\delta$ is undefined. 
Remarkably, such window turns out to be the same in the rescaled units regardless of the particle size. Note how before the reconnection the momentum of the trapped particle oscillates weakly about a constant average, because of the Magnus precession induced by the vortex \cite{giuriatoHowTrappedParticles2020a}. If the invariance (\ref{Eq:scaling}) really holds, the net particle momentum increment 
after the detachment in the rescaled units is expected to
coincide for particles of different radius but same relative mass $\mathcal{M}$. However, a small mismatch can be observed, which is probably due to the interaction between the particle and the vortex by which it is trapped before the reconnection. Such interaction indeed produces Magnus oscillations of greater amplitude for the large particle, as well as 
generation of Kelvin waves along the filament and sound radiation, which are certainly corrupting the scaling invariance (\ref{Eq:scaling}). 

We eventually analyze the momentum exchange between vortices and particle. Parametrizing the vortex ring after the reconnection as $\mathbf{R}(s,t)$, where 
$s$ is a spatial parametrization variable, the linear momentum of the vortex can be expressed within the Biot--Savart framework as \cite{pismenVorticesNonlinearFields1999}
\begin{equation}
\mathbf{p}^\mathrm{vort} = \frac{\rho_0\Gamma}{2}\oint\mathbf{R}(s,t)\times\mathrm{d}\mathbf{R}(s,t),
\label{Eq:vortex_momentum}
\end{equation}
where the contour integral is evaluated along the ring. Note that the vortex linear momentum (\ref{Eq:vortex_momentum}) is de facto a purely geometrical quantity, determined by the spatial configuration of the ring. In fact, each component of the vortex momentum can be related to the projection of the oriented area enclosed by the filament onto the corresponding direction \cite{zuccherMomentumVortexTangles2019}.
The momentum contribution of the superfluid ${\bf p}_{\rm GP}$ to the total momentum in Eq.\eqref{Eq:momcons} contains the vortex momentum \eqref{Eq:vortex_momentum} and compressible waves.  

\begin{figure}[h!]
\includegraphics[width=.99\linewidth]{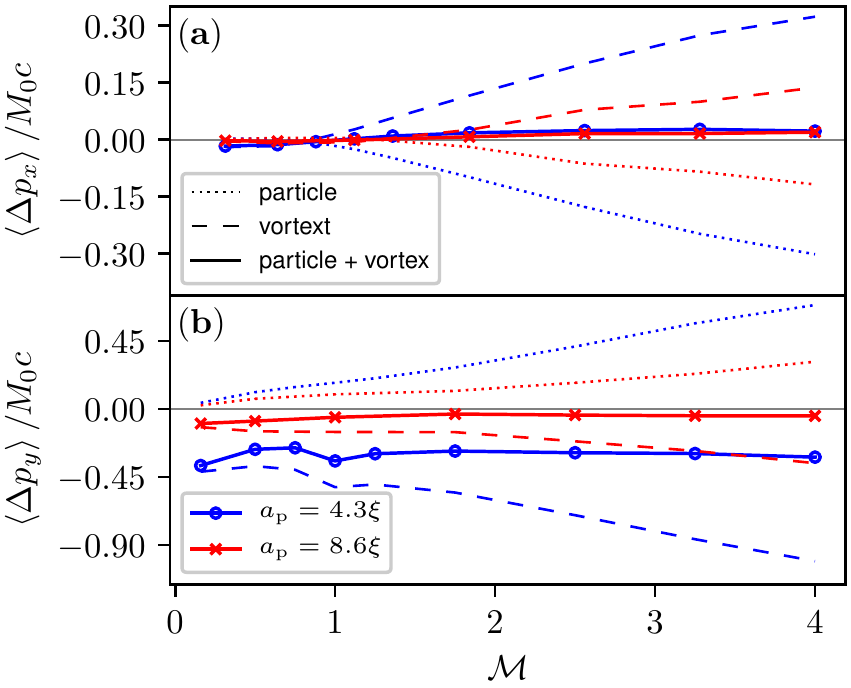}
\caption{(\textit{Color online}) $x$ component \textbf{(c)} and $y$ component \textbf{(d)} of the net momentum increment as a function of the particle mass for different particle sizes. Dotted lines are the particle momentum, dashed lines are the vortex ring momentum and solid lines the sum of the two. Blue lines refer to the small particle ($a_{\mathrm{p}}=4.3\xi$), and red lines to the large one ($a_{\mathrm{p}}=8.6\xi$).}
\label{Fig:momentum_tot}
\end{figure}

The net momentum increment for the vortex is defined as 
$\left\langle\Delta\mathbf{p}^\mathrm{vort}\right\rangle = \left\langle\mathbf{p}^\mathrm{vort}(t>t_\mathrm{rec})\right\rangle - \mathbf{p}^\mathrm{vort}(t=0)$, analogously to the net momentum increment for the particle. In practice, the vortex momentum is computed from the filaments tracked during the GP simulation. Then it is averaged over a time window of $\sim 20\tau$ after the particle detachment, in which it remains steady. The $x$ and the $y$ components of the net momentum increments as a function of the mass are displayed  Fig.\ref{Fig:momentum_tot}. 
The dotted lines are the particle net momentum increments, the dashed lines are the corresponding vortex net momentum increments and the solid lines are the sum of the two. Blue lines refer to the small particle and red lines to the large one. 

In the $x$-direction (perpendicular to the dipole velocity) the momentum acquired by the particle compensates almost exactly the momentum increment of the vortex, and thus the transfer to sound modes is negligible. On the contrary, in the $y$-direction and in particular for the small particle (solid blue line with circles), we observe a net momentum transfer from the particle and the vortices to other degrees of freedom. This transfer is independent of the particle mass and it is consistent with the observation of a sound pulse after the reconnection in Fig.\ref{Fig:dipolerec}.

\section{Reconnection of two linked rings}
\label{sec:rings}
\begin{figure*}[t!]
\includegraphics[width=.99\linewidth]{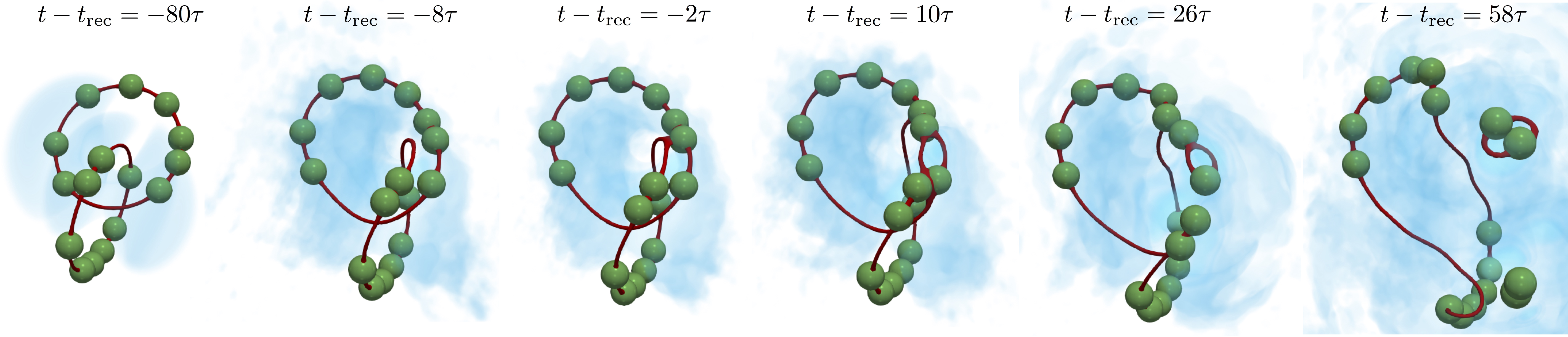}
\caption{(\textit{Color online}) Snapshots of the superfluid density and light particles ($\mathcal{M}=0.51$) during the Hopf link reconnection (time varies from left to to right). Vortices are displayed as red isosurfaces at low density, particles are the green spheres and sound is rendered in blue.}
\label{Fig:ringsrec}
\end{figure*}

In this section we study a different setting in which vortices reconnect regardless the presence of particles. In particular, we consider as initial configuration a Hopf link consisting of two vortex rings with radius $R=18\xi$, which is known to spontaneously undergo reconnection. We place $N_\mathrm{p}=8$ particles of size $a_\mathrm{p}=3.7\xi$ randomly distributed along each ring. The initial condition is shown in first snapshot on the left of Fig.\ref{Fig:ringsrec}. The numerical parameters for the particles potential are $V_0=20\mu$ and $\zeta=3\xi$.

We set as initial velocity of each particle the velocity of the ring by which it is trapped ${\bf v}_{\rm ring}$. In order to study how the precense of particles modifies the reconnection we consider three different particle masses, light ($\mathcal{M}=0.51$), neutral ($\mathcal{M}=1$) and heavy ($\mathcal{M}=3.14$ and $\mathcal{M}=12.56$). The evolution of the system for light particles ($\mathcal{M}=0.51$) according to the GP dynamics is displayed in Fig.\ref{Fig:ringsrec}.
Analogously with what observed for the dipole, as resulting of the particle-vortex interaction \cite{giuriatoInteractionActiveParticles2019}, the reconnection takes place between one trapped particle and the other filament. In the particular case of light particles, two unlinked vortex rings emerge after the reconnection: 
a large ring, which contains the majority of the particles and a small ring with two particles still attached. Moreover, because of the violence of the event, a couple of particles {gets detached from} the vortices. 

\begin{figure}[h!]
\includegraphics[width=.99\linewidth]{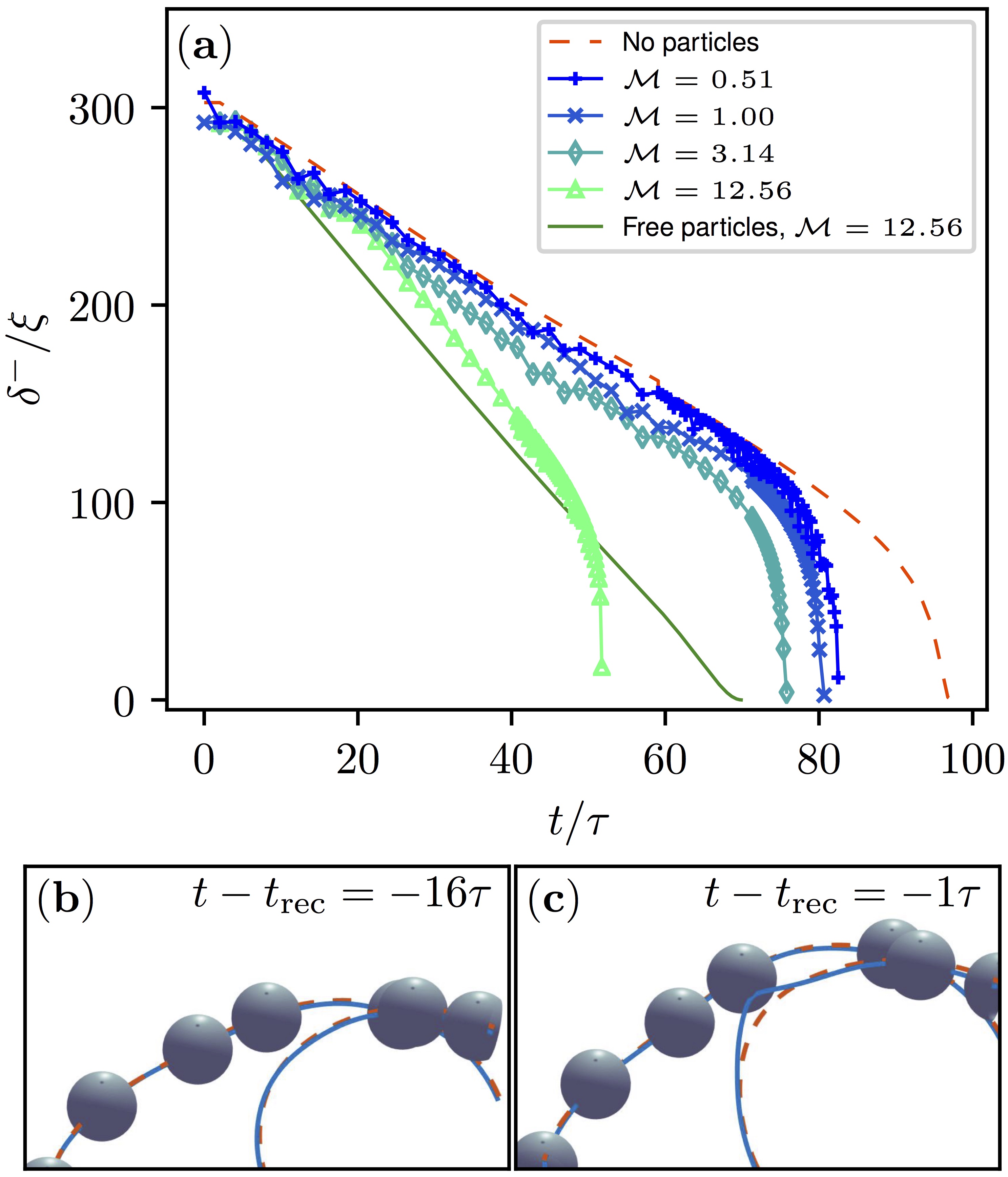}
\caption{(\textit{Color online}) \textbf{(a)} Separation between the reconnecting rings for different masses of the trapped particles (solid lines with markers). Red dashed line is the vortex separation in absence of particles and green solid line is the separation between ballistic particles without vortices. 
In the bottom panel a reconstruction of the event for light particles ($\mathcal{M}=0.51$) using the tracked filaments before \textbf{(b)} and at \textbf{(c)} the reconnection. The filaments of the simulation with particles are displayed as solid blue lines. The filaments corresponding to a simulation with the same initial condition but without particles are shown as red dashed lines.}
\label{Fig:deltaring}
\end{figure}

In order to give a quantitative description, we measured the separation rate $\delta(t)$ for the different masses. They are reported in Fig.\ref{Fig:deltaring}.a. as solid lines with markers.
For comparison, the figure also includes the distance $\delta$ for the vortices without particles (dashed red line). 
Overall, if the particle are not too heavy, the reconnection remains almost unaffected by their presence. However, at very close distance a speed up takes place due to particle-vortex interactions. Conversely, in the case of heavy particles, their inertia is so large that vortices are driven by them. To illustrate this fact, we consider the fictitious case in which free heavy particles (without vortices) are set in the same positions and with the initial velocity of the trapped ones. The distance in this case is computed as the minimal distance between the two groups of particles. Comparing this separation with the one of heavy trapped particles $\mathcal{M}=12.56$ (light green triangles), it is clear that also in this latter case the ballistic motion of particles governs the dynamics.

Finally, in the lower panel of Fig.\ref{Fig:deltaring} a reconstruction of the same event displayed in Fig.\ref{Fig:ringsrec} using the tracked vortex filaments (rendered as blue lines) is also shown, from a different perspective. For comparison, the tracked vortices 
corresponding to a simulation of the same initial configuration but without particles are shown as red lines. It is evident that the dynamics in the two cases is rather similar, especially before the reconnection. However, in the moments immediately prior the reconnection one of the vortices decorated with particles shows a clear bending towards a particle set on the other filament. This is a clear indication of a fast acceleration, which
is induced to the fluid depletion generated by the presence of the particle.  

\section{Discussion}
In this work we studied how particles trapped inside quantum vortices modify the process of vortex reconnections. We have investigated two different settings: a vortex dipole with one trapped particle and Hopf link with a number of particles randomly positioned within the vortex. Whereas in the first case the reconnection is triggered by the symmetry breaking induced by the particle, in the second one vortices reconnect regardless of the presence of particles. In the case of the dipoles, we observe that the $t^{1/2}$ temporal reconnection scaling is preserved independently of the particle mass and size. During the reconnection process, we observe a net momentum transfer from vortices to particles in both the directions perpendicular to the axis of the vortex dipole. In the transverse direction respect to the dipole initial velocity, the transfer is proportional to the mass of the particles and it is almost exactly compensated by an equal change in the vortex momentum. In the direction of the dipole displacement, the particle speed up after reconnection is not fully compensated by the vortices. The net momentum difference is roughly independent of the mass and it could be associated to the emission of a sound pulse, as the one studied in \cite{villoisIrreversibleDynamicsVortex2020}. In the case of the Hopf link vortex, it was observed that the reconnection process at large distances is almost unaffected by neutral or light particles. On the contrary, if particles are heavy it is driven by the particle ballistic motion. At very close distances, the reconnection is speeded up because of the interaction between the particles and the reconnecting vortex. 
In general, it was also observed that reconnection takes place generically between a trapped particle and an approaching filament.

In conclusion, besides providing further insights to the current knowledge of the vortex reconnection process, our findings constitute theoretical support and benchmark for the superfluid $^4\mathrm{He}$ experiments at very low temperature, in which the vortices are sampled by solid particles \cite{bewleyCharacterizationReconnectingVortices2008,paolettiReconnectionDynamicsQuantized2010}. In particular, as it has been proofed in the case of Kelvin wave tracking \cite{giuriatoHowTrappedParticles2020a},
we stress that the use of light particles is recommended for reproducing the bare vortex dynamics, provided of course that buoyancy effects remain negligible.

\appendix
\section{Dealiasing of the equations of motions and conservation of the invariants}
\label{dealiasing}
The set of equations of motion (\ref{Eq:GPEParticles},\ref{Eq:GP}) need to be dealiased in order to conserve the total momentum \eqref{Eq:momcons}. The equations are dealiased  performing a Galerkin truncation, that consists in keeping only the Fourier modes
with wavenumbers smaller than a UV cutoff $k_\mathrm{max}$. The truncated equations of motion are
\begin{eqnarray}
i\hbar\dertt{\psi} &=& \mathcal{P}_\mathrm{G}\left[- \frac{\hbar^2}{2m}\nabla^2 \psi -\mu\psi +  g\mathcal{P}_\mathrm{G}\left[|\psi|^2\right]\psi+\sum_{i=1}^{N_\mathrm{p}}\Vp^i\psi \right],
\label{Eq:GPdeal} \nonumber \\
\\
\Mp\ddot{\bf q}_i &=& - \int  \Vp^i \mathcal{P}_\mathrm{G}\left[\nabla|\psi|^2\right]\, \mathrm{d} \x+\sum_{j\neq i}^{N_\mathrm{p}}\frac{\partial}{\partial{\bf q}_i }V_\mathrm{rep}^{ij}, \nonumber \\ 
\label{Eq:Pardeal}
\end{eqnarray}
where $\Vp^i=\Vp(| \x -{\bf q}_i|)$ and  $\mathcal{P}_\mathrm{G}$  is a Galerking truncation operator. $\mathcal{P}_\mathrm{G}$ acts on a function $f(\mathbf{x})$ as 	
$\mathcal{P}_\mathrm{G}\left[f(\mathbf{x})\right]=\sum_{\mathbf{k}}\hat{f}(\mathbf{k})e^{i\mathbf{k}\cdot\mathbf{x}}\theta_\mathrm{H}(k_\mathrm{max}-|\mathbf{k}|)$, where $\hat{f}(\mathbf{k})$ is the Fourier transform of $f(\mathbf{x})$, $\theta_\mathrm{H}$ is a Heaviside theta function. It is also assumed that the particle potential is always truncated:
$V_\mathrm{p}^i=\mathcal{P}_\mathrm{G}\left[V_\mathrm{p}^i\right]$.
The equations (\ref{Eq:GPdeal},\ref{Eq:Pardeal}) exactly conserve all the invariants (Hamiltonian, fluid mass and also total momentum) if the $2/3$ rule is used, namely if $k_\mathrm{max}=\frac{2}{3}\frac{N_\mathrm{res}}{2}$, with $N_\mathrm{res}$ the number of uniform grid points per direction \cite{krstulovicEnergyCascadeSmallscale2011}. For a pseudo-spectral code, this technique implies an extra  computational cost of one extra back and forth FFTs. 

\acknowledgments{
The authors were supported by Agence Nationale de la Recherche through the project GIANTE ANR-18-CE30-0020-01. Computations were carried out on the M\'esocentre SIGAMM hosted at the Observatoire de la C\^ote d'Azur and the French HPC Cluster OCCIGEN through the GENCI allocation A0042A10385.}

%

\end{document}